\documentclass[journal]{IEEEtran}

\usepackage{lineno,hyperref}
\usepackage[utf8]{inputenc}
\usepackage{framed, color}
\usepackage{standalone}
\usepackage{tikz}
\usepackage{pgfplots}
\pgfplotsset{compat=newest}
\pgfplotsset{plot coordinates/math parser=false}
\newlength\figureheight
\newlength\figurewidth
\usepackage{float}
\usepackage{geometry}
\usepackage{booktabs}
\usepackage{tabularx}
\usepackage{easyReview}

\modulolinenumbers[5]

\begin{document}

\title{Estimation of Physiological Motion Using Highly Accelerated Continuous 2D MRI}

\author{Lynn Johann Frohwein, Florian Büther, and~Klaus Peter Schäfers
\thanks{Lynn Johann Frohwein is with the European Institute for Molecular Imaging, University of Münster, Waldeyerstra\ss e 15, 48149 Münster, Germany}%
\thanks{Florian Büther is with the Department of Nuclear Medicine, University Hospital of Münster, Albert-Schweitzer-Campus 1, 48149 Münster,
Germany}%
\thanks{Klaus Sch\"afers is with the European Institute for Molecular Imaging, University of Münster, Waldeyerstra\ss e 15, 48149 Münster, Germany}%
\thanks{L. J. F., F. B. and K. P. S. receive research grants from Siemens Healthcare GmbH on a project dealing with data-driven gating in PET/MRI.}%
\thanks{This work has been submitted to IEEE for possible publication. Copyright may be transferred without notice, after which this version may no longer be accessible.}
}

\maketitle

\begin{abstract}
Patient motion is well-known for degrading image quality during medical imaging. Especially positron emission tomography (PET) is susceptible to motion due to its usually long scan times. In hybrid PET/MRI (magnetic resonance imaging), simultaneously acquired dynamic MRI data can be used to correct for motion. Usually, MRI model-based motion correction approaches are applied to the PET data. However, these approaches may fail for non-predictable, irregular motion. We propose a novel approach for the continuous and real-time tracking of motion using highly accelerated, dynamic MRI for an accurate motion estimation. For this purpose, a TurboFLASH sequence is utilized in single-shot mode with additional exploiting GRAPPA acceleration. Sampling frequency for one slice is up to 26 ms and 520 ms for one 3D volume of 20 coronal slices. Principal component analysis and a phase-sensitive resorting of slices is performed to restore temporal consistency of the volumes. Motion is estimated from these volumes using hyper-elastic registration. The approach is validated with the help of a dynamic thorax phantom as well as with eleven healthy volunteers. Phantom ground-truth data demonstrates that the approach produces an accurate motion estimation. Volunteer validation proves that the approach is also valid for different respiratory amplitudes including highly irregular breathing. The approach could be proved to be promising for a continuous PET motion correction.
\end{abstract}



\section{Introduction}

Patient motion during long scan sessions is a well-known image degrading factor in medical imaging. In magnetic resonance imaging (MRI) the effect of motion emerges in form of geometric distortions due to incoherent k-space sampling, image blurring or ghosting artifacts. Even very small motion amplitudes can corrupt and diminish the diagnostic quality of the data \cite{Godenschweger2016}. In clinical thorax imaging, patients are therefore asked to hold their breath during image acquisition. This procedure is mandatory as in most of the cases the sampling frequency for one slice or partition of MRI data is insufficient to capture a sharp, motion-free image during free-breathing. 
However, patients are often incapable of holding their breath for this amount of time and repeatedly over a series of acquisitions.
Therefore, there is a large number of publications addressing the field of free-breathing MR for the sake of the patient's comfort, \cite{Chandarana2015,Zhang2015,Cheng2015,Zucker2018}.

A simple solution to this problem is to continuously track the motion and only acquire data during quiet episodes of breathing (triggering). However, the major disadvantage of this method is the severely prolonged scan time which, in turn, can lead to patient distress. 

In general, all methods rely on external or internal (data-driven) signals describing the motion. This motion tracking can be achieved with various approaches. Pressure-sensitive respiratory pillows can be used to measure the lift of the abdominal wall during lung expansion. However, it has been shown that the correlation of the external pillow signal and the internal organ motion is relatively low in many cases \cite{Li2017}. Furthermore, external hardware requires careful handling and potentially shifts during the scan which may corrupt the tracking. 

To find a surrogate with a high correlation to the organ motion, navigator-based motion tracking methods may be preferred. For this purpose, the investigator chooses a small region on the image data itself, e.g. liver edge for respiratory tracking. This region is repeatedly acquired and evaluated throughout the entire scan time. A 1D surrogate signal is then calculated from the intensity profiles of the navigator region. The navigator acquisition requires a special additional excitation which is interlaced between the read-outs of the anatomical sequences. This technique again prolongs the scan time. 

In the last couple of years, real-time MRI has risen to be one of the most promising new technological developments in many applications of MRI \cite{Dietz2019}, e.g. cardiovascular imaging/angiography \cite{Zhang2014}, arthrology \cite{Uecker2010,Boutin2013}, or fMRI \cite{Goncalves2017}. The method opens the possibility to not only image internal and external body movements in real-time \cite{Uecker2010}, but it can also drastically reduce scan time. Real-time MRI refers to a high sampling frequency often achieved by an undersampling of k-space. However, reconstructing a highly undersampled (sparse) k-space leads to severe imaging artifacts. Therefore, this sparse acquisition is usually combined with iterative (non-linear) inverse reconstruction algorithms dealing with the sparsity of the k-space and thereby minimizing undersampling artifacts. 

In the field of motion detection and correction of simultaneously acquired positron emission tomography (PET) data in hybrid PET/MRI, most methods utilize a 3D approach of the described real-time MRI acquisition and reconstruction schemes. For instance, Block et al. \cite{Block2014} and Grimm et al. \cite{Grimm2015} use a golden-angle radial stack-of-stars sampling scheme with phase-encoding in z-direction, whereas Kolbitsch et al. follow a very similar approach with a 3D golden radial phase encoding (GRPE) in the x-y-plane \cite{Kolbitsch2014,Kolbitsch2017,Kolbitsch2019}. A respiratory surrogate signal is derived through analysis of the k-space center (\emph{self-gating}) which is acquired in every volume repetition. Using this signal, the k-space lines (spokes) can be sorted into different gates belonging to different respiratory phases (e.g. maximum expiration to maximum inspiration). These gates can be reconstructed individually and used for the creation of motion vector fields which can be used to motion-correct the PET data.

The major limitation of these methods is that it is either necessary to acquire many spokes to be able to reconstruct an image with few artifacts or to reconstruct the data with computationally expensive compressed-sensing algorithms.
Most approaches for the motion correction of PET data in hybrid PET/MRI use model-based schemes \cite{McClelland2013}. In these schemes, MR-based motion data is acquired only for a short time section of the entire PET scan time to allow other diagnostic sequences during one PET acquisition. Subsequently, this motion data is used to correct for motion of the entire scan time. These approaches assume a very periodic motion pattern. Changing breathing patterns, irregular motion or baseline changes during the PET scan can not be accounted for correctly. However, the unique feature of hybrid PET/MRI scanners is the simultaneous acquisition of PET and MRI data. Therefore, MR-based motion data can be derived throughout the whole PET scan duration to achieve an exact motion correction. 

The main goal of this study is the demonstration of a novel approach for continuous and real-time motion detection with a fast 2D MRI sequence providing precise spatio-temporal motion vector fields for whole-body PET/MRI. The present work is part of an exact motion correction approach for whole-body PET data in hybrid PET/MRI, while this study focuses on the MR-based motion estimation.

\section{Methods \& Material}

\subsection{Data acquisition}
All acquisitions in this study were performed on a Siemens Biograph mMR (Siemens Healthcare GmbH, Erlangen, Germany) 3T hybrid PET/MRI. 
The sequence used for the acquisition is a cartesian TurboFLASH sequence. For a fast acquisition, we used the single-shot mode of the sequence and exploited the parallel acquisition technique (PAT) functionality of the scanner. To enable a large acceleration factor of PAT=8 while maintaining the signal-to-noise ratio (SNR), we utilized a 32-channel phased-array cardiac coil (In Vivo Corporation, Gainesville, FL, USA) consisting of an anterior and a posterior part.

\begin{figure*}
    \centering
    \includegraphics[width=0.8\textwidth]{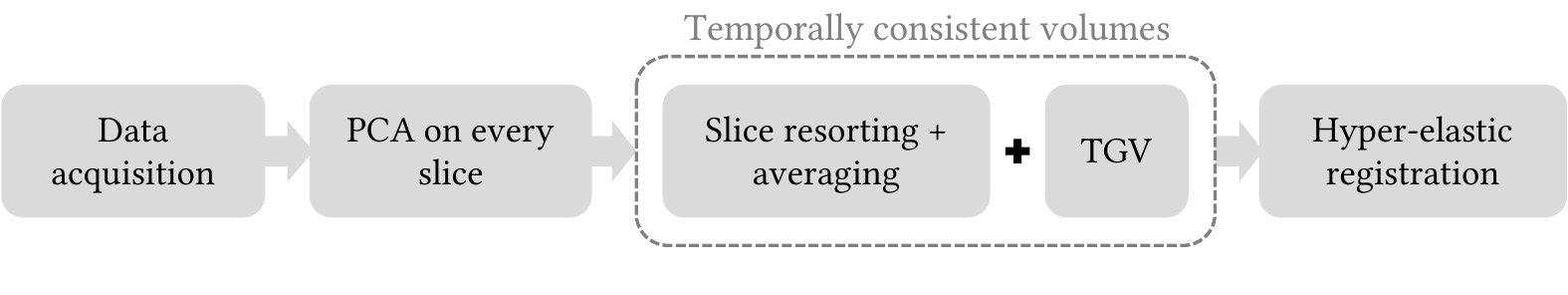}
    \caption{Workflow of the post-processing. After data acquisition, slices are resorted with the help of PCA-based surrogate signals to create temporally consistent volumes. To improve SNR, slices showing the same resp. amplitude are averaged before residual noise is remove with TGV. Motion is then estimated with hyper-elastic registration.}
    \label{fig:flow}
\end{figure*}

\begin{figure}
    \centering
    \includegraphics[width=0.99\columnwidth]{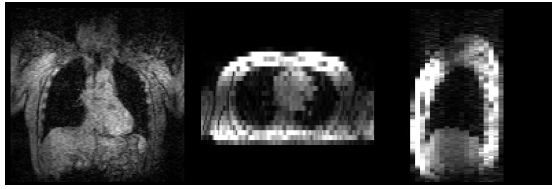}
    \caption{Volunteer data acquired with the TurboFLASH sequence: (left): sagittal slice through heart, (middle): transversal central slice, (right): sagittal slice though liver. Temporal resolution was 35~ms/slice, 700 ~ms/volume. The temporal offset between even and odd slices is clearly visible in the form of a stepped liver edge. 
    }
    \label{fig:example}
\end{figure}

This study mainly focuses on thoracic motion. Therefore, the coil was approximately centered at the sternum to image large parts of the lungs, the heart and the liver. Figure \ref{fig:example} shows an example of the image data that was acquired with the sequence.
For all scans, 20 coronal slices per volume were acquired with an in-plane spatial resolution between 3.1 x 3.1~mm$^2$ and 3.9 x 3.9~mm$^2$ (FoV of 400-500 $mm^2$), a slice thickness of 6.0-7.0~mm, $T_R$ = 26-35~ms, $T_E$ = 1.3~ms, bandwidth = 1953 Hz/px and an acceleration factor of PAT=8 using GRAPPA parallelization \cite{Griswold2002}. 
Due to the single-shot functionality of the sequence, one slice is acquired during one $T_R$. The sampling frequency for one slice was up to 38~Hz and up to 1.9~Hz for a volume of 20 slices, respectively leading to an acquisition time of 520-700~ms per volume. A scan time of 10 minutes resulted in 900-1000 3D volumes.
Moreover, slices were acquired in interleaved mode. In this mode, all even slice positions were acquired before the odd slice numbers. This technique is usually utilized to avoid cross-talk between adjacent slices. Here, we used this scheme to sample almost the entire volume extent twice in one volume repetition. This way, the central parts of the body showing the most prominent motion, namely the liver dome and the heart, were sampled regularly with only half of the full-volume repetition time.

\subsection{Volume creation and motion estimation}
The post-processing workflow is depicted in Figure \ref{fig:flow}. All steps were automatized and implemented in \emph{MATLAB} (The MathWorks, Inc.).
As data is acquired in 2D, there is always a slight temporal offset between the adjacent slices of a volume (see Figure \ref{fig:example}) which has to be corrected to avoid errors in the motion estimation. For this purpose, we used respiratory surrogates for every slice position.
\bigskip
\subsubsection{Respiratory surrogate via PCA}
The first step in the post-processing was the determination of respiratory surrogate signals. These were determined using the principal-component analysis (PCA). Similar techniques using PCA have been utilized before for the determination of respiratory or cardiac traces in various modalities (e.g. \cite{Thielemans2011,Gupta2016,Hamy2014}). 

The first principal-component represents the component with the largest variance. In a dynamic image series, this usually refers to respiration. The advantage of using PCA for the determination of surrogate signals is the automatability as neither a navigator has to be selected manually on the image data nor additional hardware (e.g. respiratory pillow/belt, tracking devices, etc.) is needed. Furthermore, the first principal-component of the PCA by itself is independent of noise. 

\begin{figure}
    \centering
    \includegraphics[width=0.99\columnwidth]{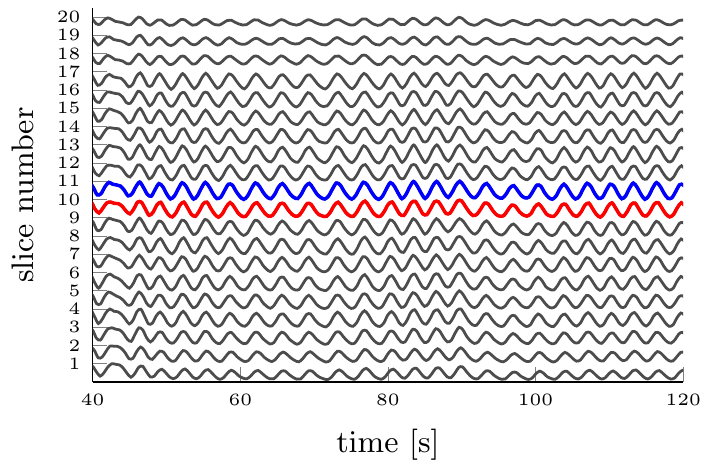}
    \caption{PCA-based signals (weights of 1$^{st}$ component) of every slice position (top). The collection of signals shows that the respiratory signal could be reliably derived from every slice position. Slices 18-20 show a lower amplitude as they are slice positions on the posterior edge of the FoV showing the back region of the volunteer. The blue and red signals are central slices used as a common surrogate for the resorting.}
    \label{fig:concat}
     
\end{figure}

The PCA was performed on each of the 20 slice positions over time. The weights of the first component were extracted and used as respiratory surrogate. These surrogate curves each have a temporal resolution of 520-700~ms (native volume repetition time). To further increase the temporal resolution, the time points of two adjacent central slices (e.g. slice 10 and 11, being half of the volume repetition time apart) were interlaced to a single signal. A high cross-correlation coefficient of R$>$0.98 between the two signals ensured accordance of both curves. Figure \ref{fig:concat} shows the extracted first PCA weights for every slice position an exemplary data set. Note that even in the most lateral slice positions (1, 2, 3 and 18, 19, 20) a reliable PCA signal could be derived. 
\bigskip
\subsubsection{Temporally consistent volumes}
\begin{figure*}
    \centering
    \includegraphics[width=0.99\textwidth]{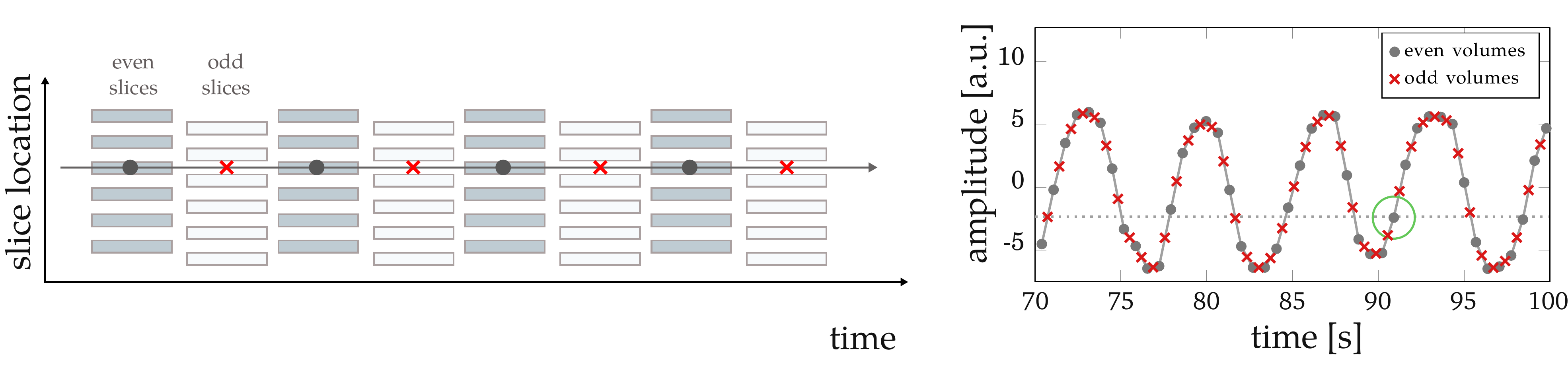}
    \caption{Slice resorting scheme: data was acquired in interleaved order. Even and odd slices were treated as separate sub-volumes. Every sub-volume (red cross) was supplemented (green circle) with slices from the other sub-volume (gray dot) with respect to the PCA-based respiratory curve. This process was conducted with every slice position. This way, the number of volumes was doubled (artificial doubling of the sampling frequency) while simultaneously respecting the real motion state.}
    \label{fig:slicesort}
\end{figure*}

To finally achieve temporally consistent volumes, the slices have to be resorted. As all slices have different sampling time points, the time points of the common surrogate signal were used as reference. At each time point of the common signal, amplitudes on each of the 20 slice positions were interpolated linearly. As the signals were densely sampled, errors due to interpolation were negligible. Matching slices for every time point were searched within all available slices. Ten slices with the smallest euclidean distance in terms of respiratory amplitude were averaged (respiratory phase-sensitive averaging) and sorted into the new volume. We chose this number of slices as a balance between improvement of SNR and phase conformity. In all analyzed cases, the mean deviation (euclidean distance) of the found slices to the interpolated amplitude was smaller than 2\% (of the maximum amplitude of PCA-based signal) in a 5-minutes scan. This deviation was not visible in the image data and was thus considered to be sufficiently accurate for the averaging. Using this scheme, the temporal intra-volume offset was eliminated while the temporal resolution of the volumes was artificially doubled and SNR was improved. Figure \ref{fig:slicesort} demonstrates the volume resorting.
SNR was measured by the following equation \cite{Kaufman1989}:

\begin{equation}
    SNR = 0.655 \cdot \frac{S_{liver}}{\sigma_{noise}}
\end{equation}

where $S_{liver}$ is the average intensity inside of a ROI in a homogeneous region, in this case the liver, and $\sigma_{noise}$ depicts the standard deviation of intensity values inside of a ROI placed in noisy regions of the image outside of the anatomy. The constant factor of 0.655 corrects for the Rician noise characteristics of the MRI images, as proposed in \cite{Kaufman1989}.

\bigskip
\subsubsection{TGV image denoising}
As the data was acquired with a very high acceleration factor, SNR was potentially low by $SNR \propto \frac{1}{\sqrt{PAT}}$ with acceleration factor PAT \cite{Breuer2009}. This may be detrimental to the calculation of the vector fields. Therefore, noise in the image data was removed with an iterative total generalized variation (TGV) denoising approach. 
%
For detailed information on this method, refer to \cite{Knoll2011,Bredies2010}.

With this method, data was denoised while simultaneously preserving edges that are important for an exact motion detection. Prior to this study, we tested several parameter sets and picked the following for all data sets according to visual improvement of the image data: $\alpha_1 = 1$ and $\alpha_0 = 0.05$ (see \cite{Knoll2011}). It could be proved to be valid for all data sets acquired with the same acquisition parameters. 
\bigskip
\subsubsection{Motion vector fields}
In a last step, motion vector fields were determined using a hyper-elastic multi-level image registration provided by the FAIR toolbox \cite{FAIR}. The hyper-elastic regularization expands the linear elastic to a non-linear model leading to more reliable registration results \cite{Burger2013,Gigengack2012}. The hyper-elastic regularization individually controls length, area and volume of the transformation through weighting parameters $\alpha_{length}$, $\alpha_{area}$ and $\alpha_{volume}$. For a detailed explanation of the method see \cite{Burger2013,Gigengack2012}.

Here, we used 7 resolution levels and the following hyper-elastic parameters: $\alpha_{length} = 10$, $\alpha_{area} = 0.1$, $\alpha_{volume} = 0.1$. The computational time for the registration of one volume was less than a minute on a single CPU core (Intel Xeon CPU @ 2.30GHz).

Each volume was registered to a reference volume which was chosen - in the context of hybrid PET/MRI - with respect to the attenuation map acquired for the attenuation correction of the PET data. To this end, the correlation between all MRI volumes and the standard DIXON water image was determined to find the correct respiratory phase of the attenuation map. The best correlating volume was chosen as the reference volume.

\begin{figure}
    \centering
    \includegraphics[width=\columnwidth]{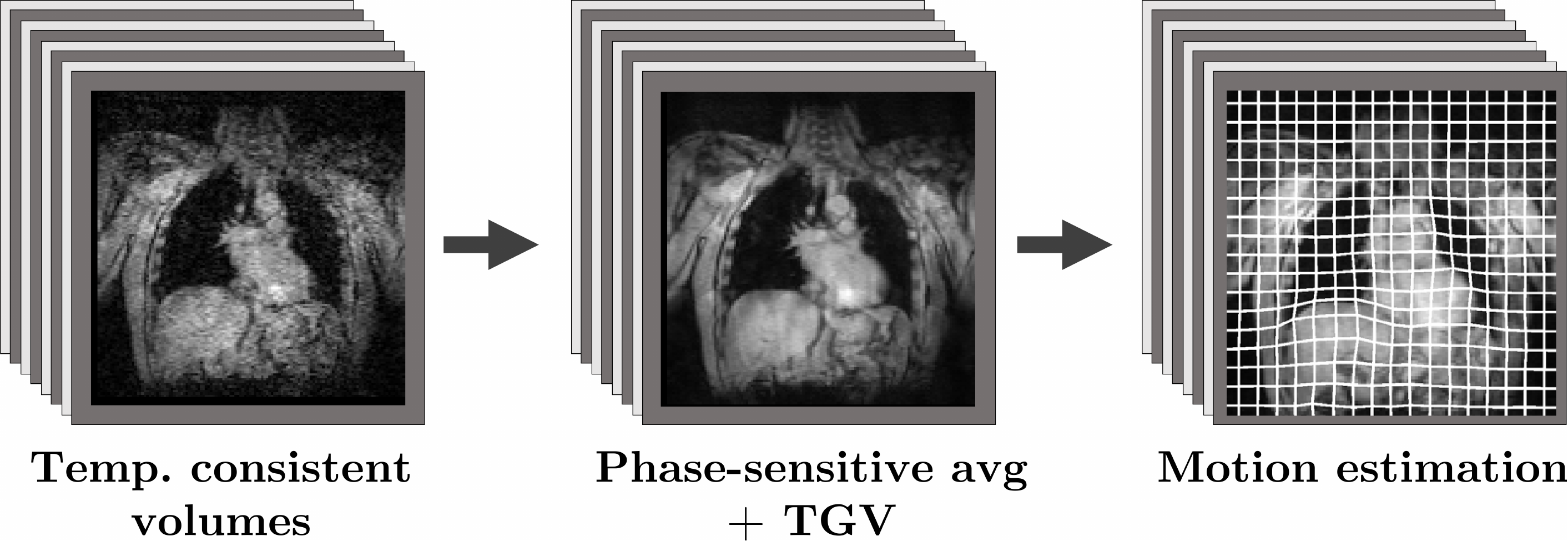}
    \caption{After resorting, the SNR was increased by averaging ten of the best phase-matching slices (phase-sensitive averaging) and additionally denoised using TGV. After this process, motion was estimated using hyper-elastic registration.}
    \label{fig:workflow}
\end{figure}

\subsection{Phantom validation}
To validate the method, MR data of a dynamic MR-compatible anthropomorphic thorax phantom 
described in \cite{Bolwin2018} was acquired. The phantom consists of a thorax with inflatable lungs, moving and beating heart (left ventricle), a moving diaphragm, a liver compartment, and a small fillable lesion. A maximum respiratory lift (lesion displacement) was pre-set to 20~mm for the phantom acquisition. 
%
TurboFLASH MRI sequences were acquired under respiratory and cardiac motion for 4.5 minutes with a $T_R$~=~26~ms, 3.9 x 3.9 x 7.0 mm$^3$ voxel size at a matrix size of 128 x 128 x 20 px.

Additionally, the phantom was scanned without motion in end-inspiration and end-expiration phase. The respiratory cycle was set to 0.14~Hz (3~s inspiration, 4~s expiration). The frequency of the respiration was modified with a slight variation so that the signal shows a more realistic pattern. The variation was achieved by randomly varying the opening and closing time of the pneumatic valves of the phantom. For all dynamic experiments, the sensor signals of the respiration were recorded in log files. 

In the analysis of the data, the resorting method was validated by a geometric comparison of the phantom's liver edge acquired in static mode. The motion detection was evaluated with the help of the sensor signals of the phantom. Furthermore, calculated motion vectors were analyzed and compared to the known motion of the phantom.

\subsection{Volunteer experiment}
In addition to the phantom experiment, we tested the MRI sequence on 10 healthy volunteers. All volunteers provided written consent for the scans. Scan duration in these scans was between 1 and 10 minutes. No breathing instructions were given to the volunteers to mimic realistic free-breathing imaging.
One additional volunteer was asked to breath with varying patterns which should include deep breathes. In all cases, we compared the determined motion vectors in the liver area to a manually determined maximum displacement of the liver edge.

\section{Results}

\subsection{Phantom validation}
\begin{figure}
    \centering
    \includegraphics[width=0.5\textwidth]{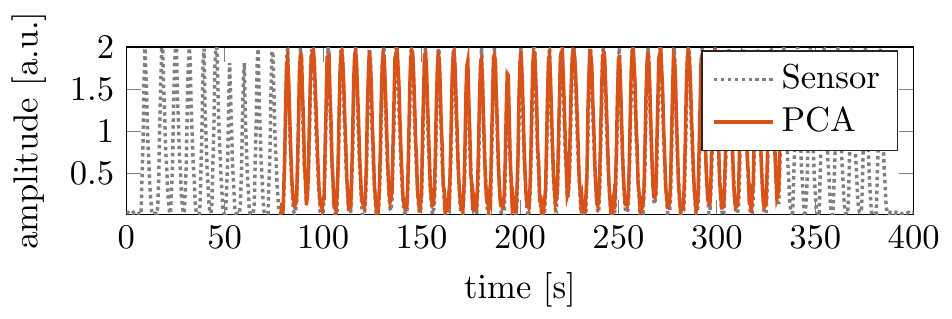}
    \vspace{-0.5cm}
    \caption{Phantom sensor signal (ground-truth) vs. weights of $1^{st}$ PCA component. The PCA signal shows a good accordance to the signal from the analog read-out of the phantom. The applied ground-truth frequency of 0.14 Hz can be identified with the PCA signal. The Pearson correlation coefficient of both signals is determined to be r$>$0.98.}
    \label{fig:PCAwilli}
\end{figure}

Figure \ref{fig:PCAwilli} displays the correlation of the weights of the first principal-component and the actual analog signal from the phantom. The Pearson correlation coefficient was determined to be r$>$0.98 for the two curves. Furthermore, the variation applied to the pneumatic system can clearly be observed in the PCA signal.  

\begin{figure}
\centering
\includegraphics[width=.99\columnwidth]{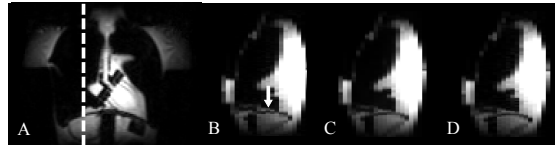}
\caption{Result of the resorting. (A): coronal slice indicating the slice position of the sagittal slice, (B): liver edge with visible temporal offset between slices manifesting as stepped contour (arrow). The offset results from the slightly different acquisition times of the slices, (C): the liver edge contour is completely restored, compared to the static reference (D)}
\label{fig:resortingresult}
\end{figure}

A slight temporal offset between even and odd slices could be observed at the liver edge in sagittal image direction (Figure \ref{fig:resortingresult}). After the slice sorting and the phase-sensitive averaging using TGV denoising the offset was eliminated. The shape of the liver edge now matches the shape extracted from a static scan in every motion phase. Pearson correlation between the reference and the resorted volume was increased from r=0.83 to r=0.99. The SNR was improved through the averaging and additional TGV denoising from 10.8 to 38.6.

\begin{figure}
    \centering
    \includegraphics[width=0.5\textwidth]{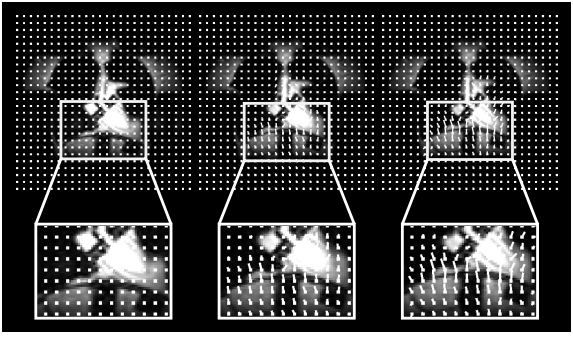}
    \caption{Vector fields for the phantom data: end-inspiration (left) as reference with no motion, mid-expiration (middle) and end-expiration (right) showing largest motion vectors. The vectors in areas of the liver are pointing in cranial direction which is in accordance with the phantom mechanics. The heart area shows a cranio-lateral motion. This motion results from the diaphragm motion pushing the heart compartment.}
    \label{fig:willigrid}
\end{figure}

Motion vector fields determined with these new volumes using hyper-elastic registration showed plausible properties regarding vector amplitudes and angles in different regions. Figure \ref{fig:willigrid} depicts three respiratory phases and the respective motion vector fields. For the vector field from end-expiration to end-inspiration (maximum displacement of 20~mm pre-set) the mean vector length for the liver area was determined to be 19.1~mm. The rigid up-and-down movement of the liver compartment was also reflected by the motion vectors. The phantom's heart is pushed in cranio-lateral direction. This motion could also clearly be observed in the motion vectors.

\subsection{Volunteer experiments}
\begin{figure}[ht]
    \centering
    \includegraphics[width=0.99\columnwidth]{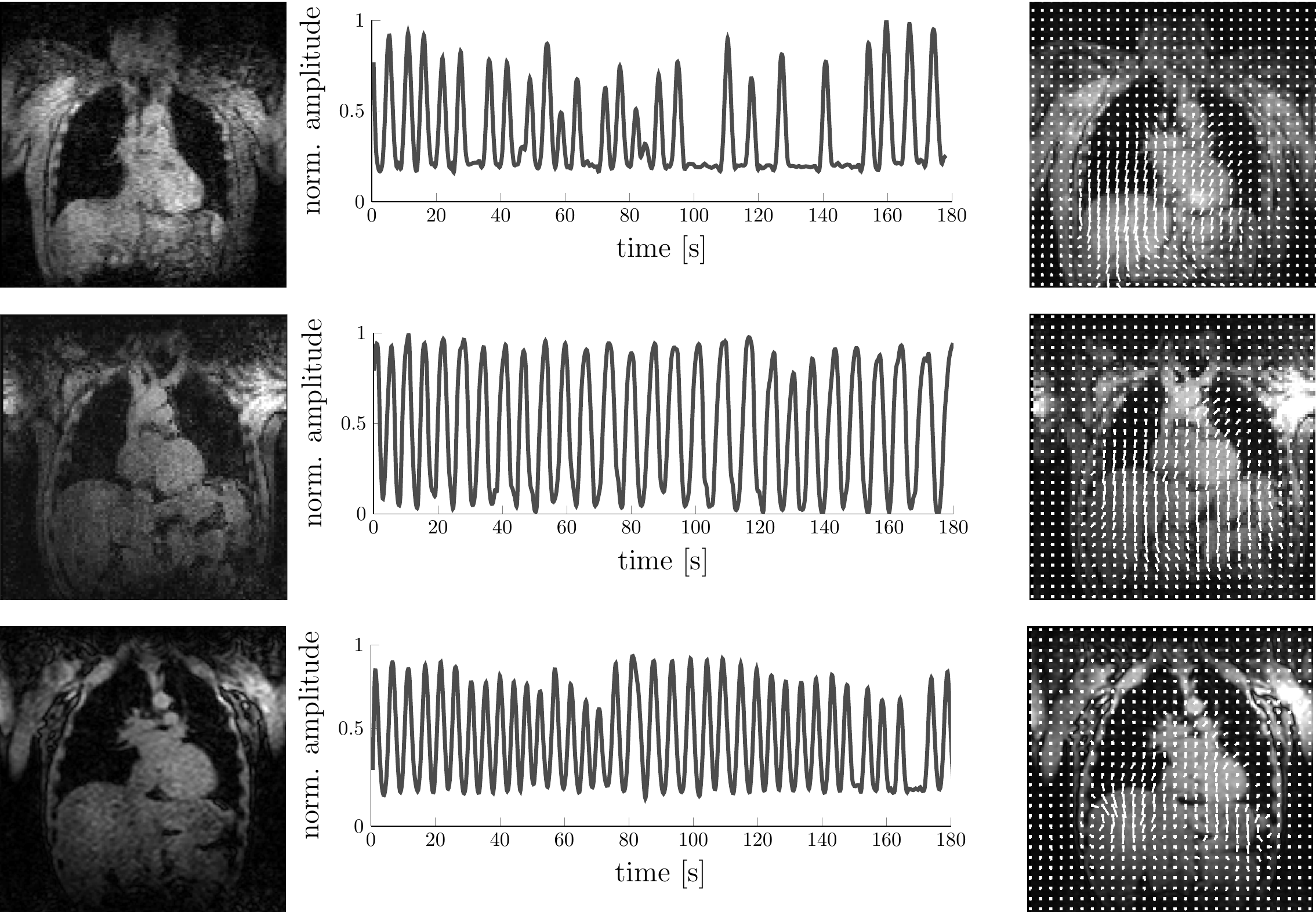}
    \caption{Three examples of volunteer data (see Table \ref{tab:volunteers}). Original image data (left), PCA-based respiratory trace (middle), corresponding motion vector field end-expiration to end-inspiration (right). The respiratory traces shown here were picked from the first principal-component of slice number 10. This slice position is centered inside of the image volume and usually displays the liver dome. The sampling frequency was 700 ms which corresponded to the volume repetition time. Volunteer 1 (top row) shows several episodes of shallow breathing with durations between 5 and 10 s as well as shifting respiratory amplitudes.}
    \label{fig:volunteersignals}
\end{figure}

A comparison of manually determined liver displacement and analysis of the motion vector fields is depicted in Table \ref{tab:volunteers}.
Figure \ref{fig:volunteersignals} shows image data of volunteers and their corresponding respiratory trace determined by PCA. Sampling frequency for all three volunteers was 700 ms (volume repetition). Especially volunteer 1 (top row of Figure \ref{fig:volunteersignals}) showed a relatively irregular breathing pattern with shallow phases of several seconds and changing amplitudes.  

\begin{table}[t]
\caption{List of volunteers: Avg. respiratory frequency, avg. resp. amplitude in liver area as measured manually and determined from the motion vector}
\resizebox{0.5\textwidth}{!}{
\begin{tabular}{cccc}
\hline
\hline
& & & \\
\textbf{Volunteers} & \textbf{avg. resp. freq. [Hz]} & \multicolumn{2}{c}{\textbf{avg. resp. amplitude [mm]}}   \\ 
\cmidrule(lr){3-4}
           &            & manual           & MVF             \\\hline
1          &    0.19    &     21.9           &      21.4        \\ 
2          &    0.20    &     17.9           &      17.5        \\ 
3          &    0.17    &     16.1           &      15.3        \\ 
4          &    0.25    &     14.0           &      13.3        \\
5          &    0.24    &     19.5           &      18.7        \\ 
6          &    0.09    &     10.8           &      10.1        \\ 
7          &    0.19    &     20.3           &      19.2        \\ 
8          &    0.13    &     12.5           &      11.8        \\ 
9          &    0.27    &     7.8            &      7.5         \\ 
10         &    0.26    &     15.6           &      15.2        \\ 

\hline
\end{tabular}
}
\label{tab:volunteers}
\end{table}

\begin{figure*}
\begin{minipage}[t][70mm]{0.99\textwidth}
    \centering
    \includegraphics[width=0.8\textwidth]{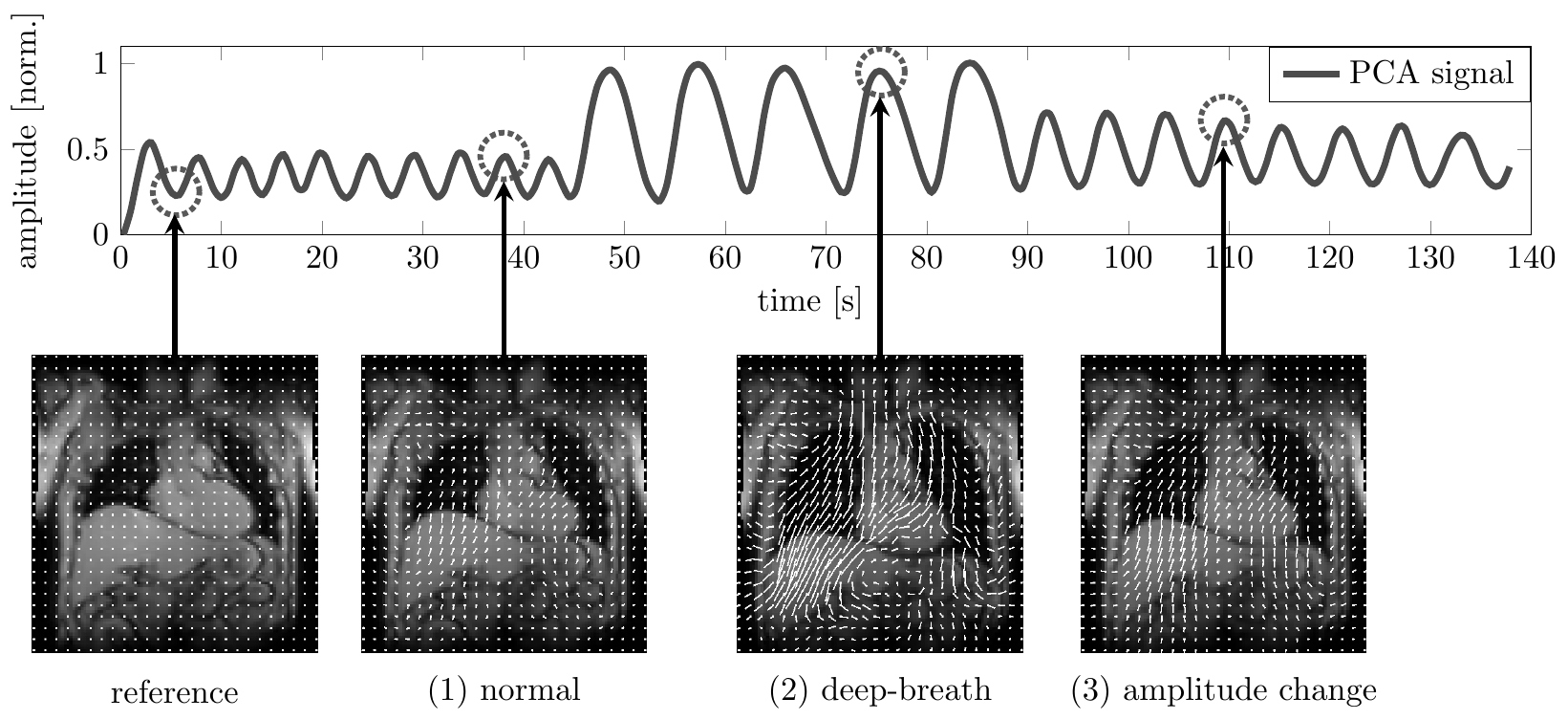}
\end{minipage}
\begin{minipage}[t][22mm]{0.99\textwidth}
    \centering
    \vspace{-1cm}
    \resizebox{0.5\textwidth}{!}{
    \begin{tabular}{llll}
    \hline\hline
    \textbf{Action} & \textbf{Vector lengths} & \textbf{Ground-truth} \\
    \hline
    & & & \\
    \textbf{(1) normal} & 13.6 $\pm$ 0.73 mm & 14.0 mm \\
    \textbf{(2) deep-breath} & 34.7 $\pm$ 3.55 mm & 35.0 mm \\
    \textbf{(3) amplitude change} & 17.1 $\pm$ 1.05 mm & 17.5 mm \\
    \hline
    \end{tabular}}
\end{minipage}
\vspace{-1cm}
    \caption{Volunteer scan with changing respiratory trace. Top: PCA signal showing respiratory trace. 1/3 of the time the volunteer showed relatively shallow breathing pattern which changed to deep breathing with large respiratory amplitudes, followed by a shallow breathing pattern with slightly larger amplitude compared to the first part. Vector fields showed good accordance to the respiratory trace. Bottom: Vector lengths at liver edge compared to manually determined maximum amplitude of liver edge.}
    \label{fig:deepbreath}
\end{figure*}

Figure \ref{fig:deepbreath} illustrates a volunteer scan with irregular breathing patterns. The first third of the scan time showed normal breathing. In the second third the volunteer was asked to breath deeply for five cycles and then return to normal breathing. The last third of the scan showed normal breathing but with altered amplitudes and a slight baseline shift. Respective vector fields corresponded well to the respiratory trace and a manual determination of respiratory amplitudes. 

\section{Discussion}

The MRI TurboFLASH sequence used within this study has been proved to be capable of acquiring data that can be used for 3D motion detection. Its ability to acquire sharp and undistorted images could be shown for phantom and volunteer data. The unique feature of simultaneous PET and MRI acquisition can be fully exploited to achieve an exact motion detection with the help of the presented MR-based methodology. This may be in particular important for future generations of PET scanners with increased spatial resolution where a precise motion vector field may provide resolution enhancements. This is in line with Polycarpou et al. who state that motion correction is mandatory to benefit from high-resolution PET systems \cite{Polycarpou2014}. Thus, it is conceivable that an increasingly accurate motion correction improves lesion detection accuracy in high-resolution PET scanners.

However, due to the short $T_R$ the images show a lack of contrast and SNR. In terms of contrast, a radial k-space sampling is beneficial over a cartesian. In principal, the sequence can easily be modified towards a radial sampling. Nevertheless, for our purpose of tracking respiratory and cardiac motion the contrast is sufficient as these types of motion manifest with relatively high amplitudes over large areas of the image (liver, lung, heart). Low contrast properties of larger tissue areas could be even beneficial for the hyper-elastic image registration due to an equal distribution of intensity values. 

At the moment, the sequence is implemented as a separate scan in the imaging workflow. However, due to the low acquisition time of around 30~ms per image, it is easily conceivable that it can be implemented as a 2D navigator-sequence interlaced between the excitations of the clinical sequences, similar to the approach in \cite{Rigie2019}. However, one limiting factor could be that clinical sequences have to be modified to incorporate this navigator acquisition. 

The sequence in combination with the proposed scheme can also be used for the creation of a MR-based motion model for the usage in hybrid PET/MRI \cite{McClelland2013}. In a pre-trial we investigated how much data has to be acquired to be able to create 10 respiratory gates which can be used for the motion estimation. The overall scan time for the acquisition of sufficient motion data is determined to be around 30~s. This small amount of time does not substantially increase the overall scan time, thus does not increase patient's distress. In combination with a respiratory signal acquired throughout the entire PET scan time (e.g. with external hardware, data-driven methods, etc.), a motion model can be created and used for the PET motion correction. 

For the proposed scheme, we used 20 slices and the vendor-provided interleaved acquisition scheme with two slice-groups (even and odd slice numbers). However, expanding the scheme to three or more slice groups could provide a super-resolution approach, in which the signal is derived from the sub-volumes which are then resorted and stitched together to create high-resolution volumes. This could further increase imaging speed or enable more slices increasing the spatial resolution.

In our approach, we used a PCA-based respiratory signal derivation. We have chosen this technique because the computational time for an entire time series of 5000 3D volumes was measured to be 9 seconds. We also tested an Independent Component Analysis (ICA) to derive the signals. However, the computational time for the same data set was with 70 minutes substantially longer and could not yield a crucial benefit over the very fast PCA signal derivation. It still has to be further investigated whether an alternative approach could also detect the cardiac signal. As shown in \cite{Rigie2019}, second‐order blind identification (SOBI) algorithm can successfully separate respiratory and cardiac signals. 

As already mentioned in the introduction, compressed sensing approaches can largely improve noisy and artifact-corrupted data. These approaches can also be applied to Cartesian MRI data. However, the sampling of the k-space data has to be modified to justify a pseudo-random sparsity necessary for compressed sensing. The focus of this study is on the post-processing of data that has already been reconstructed on the scanner. In a future study, we will also include reconstruction of the data into the workflow integrating most of the post-processing steps presented herein. 

Furthermore, there are already numerous publications dealing with denoising of medical images (or raw k-space in case of MRI) with the help of machine learning, especially with neural networks, e.g. \cite{Kaur2018,Jiang2018}. The technology is promising, as neural networks can be trained with noisy and noise-free images to learn the noise distribution of the data presented. This way, the data can be highly undersampled and still be reconstructed with reasonable diagnostic quality.

\section{Conclusion}

In this study, we have developed a methodology for generating continuous three-dimensional whole-body MRI data which allows precise calculation of motion vector fields that can be used for PET motion correction in hybrid PET/MRI. The applied MRI sequence set-up proposed herein has the ability of acquiring sharp, undistorted images using very short slice acquisition times. The post-processing has been demonstrated to be capable of reliably and automatically track respiratory motion via PCA in periodically as well as irregular motion patterns which can be utilized to generate continuous image volumes. Validation of the approach using a dynamic anthropomorphic thorax phantom renders promising results for an exact, continuous motion correction in simultaneous PET/MRI.

\section*{Acknowledgements}
This study was partly supported by the Deutsche Forschungsgemeinschaft (SFB 656), and the cluster of excellence Cells-in-Motion (EXC 1003 – CiM). The authors kindly thank the technician team Anne Kanzog and Stanislav Milachowski for their technical support and Björn Czekalla for his support regarding phantom preparation.

\bibliography{bibliography}

\end{document}